\documentclass[journal,twoside,web]{ieeecolor}
\usepackage{generic}
\usepackage{cite}
\usepackage{amsmath,amssymb,amsfonts}
\usepackage{algorithmic}
\usepackage{graphicx}
\usepackage{algorithm,algorithmic}
\usepackage{hyperref}
\usepackage{wrapfig}
\usepackage{tabularx}
\usepackage{caption}
\usepackage{subcaption}
\hypersetup{hidelinks=true}
\usepackage{textcomp}
\usepackage{multirow}
\usepackage[export]{adjustbox}
\usepackage[table]{xcolor}
\def\BibTeX{{\rm B\kern-.05em{\sc i\kern-.025em b}\kern-.08em
    T\kern-.1667em\lower.7ex\hbox{E}\kern-.125emX}}
\usepackage[utf8]{inputenc}
\usepackage{textgreek}
\DeclareUnicodeCharacter{03B2}{\textbeta}
\begin{document}
\title{Gait Analysis using 6DoF Magnetic Tracking}
\author{R. Abhishek Shankar, Hyungjun Ha, and Byunghoo Jung, \IEEEmembership{Member, IEEE}
\thanks{Manuscript received 22nd April, 2024. \textit{(Corresponding author: R. Abhishek Shankar.)} }
\thanks{Hyungjun Ha, R. Abhishek Shankar, and Byunghoo Jung are associated with the Department of Electrical and Computer Engineering at Purdue University, West Lafayette, IN 47906 USA (e-mail: shanka18@purdue.edu; ha101@purdue.edu; jungb@purdue.edu) }}

\maketitle

\begin{abstract}
Gait analysis using wearable devices

Gait analysis using wearable devices has advantages over non-wearable devices when it comes to portability and accessibility. However, non-wearable devices have consistently shown superior performance in terms of the gait information they can provide. This calls for the need to improve the performance of wearable device based gait analysis. To that end, we developed a 6 Degrees-of-Freedom (6DoF) magnetic tracking based gait analysis system as a step in this direction. The system is portable, minimally intrusive, wireless and power efficient. As a proof-of-concept, the system was used for the task of Human Activity Recognition (HAR) to classify four tasks - walking (W), walking with weight (WW), jogging (J) and marching on the spot (M). Gait data of 12 participants was collected. The classification performance of two deep learning (DL) classifiers - Convolutional Neural Networks (CNN) and Long Short Term Memory (LSTM) - was compared. The performance of the magnetic tracking based gait analysis system was also compared with an Inertial Measurement Unit (IMU) + magnetometer based system. The magnetic tracking based system showed an overall classification accuracy of 92\% compared to 86.69\% for the IMU + magnetometer system. Moreover, the magnetic tracking system showed an improvement of about 8\% in being able to differentiate between W and WW. This highlights the insufficiency in the information content in the data from IMU + magnetometer, warranting the need for a complete 6DoF tracking. Our work, thus, proves the feasibility of using magnetic tracking systems for the purpose of gait analysis.
\end{abstract}

\begin{IEEEkeywords}
Human gait analysis, magnetic tracking, Human Activity Recognition, SVM, K-means clustering, PCA
\end{IEEEkeywords}

\section{Introduction}
\label{sec:introduction}
\IEEEPARstart{H}{uman} gait is a complex motor skill, that potentially contains information about the other aspects of individual human actions, behaviour and physiology. As a result, there has been a growing interest in using gait for a variety of applications, such as, biomedical devices \cite{beauchet_poor_2016}\cite{cicirelli_human_2022}\cite{abdulhay_gait_2018}\cite{rehman_comparison_2019}, sports training \cite{tabrizi_comparative_2020}\cite{stoeve_laboratory_2021}\cite{rum_wearable_2021}, Human Computer Interaction (HCI) \cite{zhang_deep_2020}, biometrics \cite{terrier_gait_2020}\cite{nambiar_gait-based_2019}\cite{filipi_goncalves_dos_santos_gait_2022} and Human Activity Recognition (HAR) \cite{murad_deep_2017}\cite{golestani_human_2020}. This drive has been accelerated by the innovations in sensor technologies, miniaturization, classification methods and embedded compute power. 

Medical research has found linkage between neurological ailments and their effect on gait \cite{beauchet_poor_2016}\cite{montero-odasso_association_2017}\cite{nadkarni_brain_2017}\cite{mirelman_gait_2019}. For example, the onset of Alzheimer's disease causes hyperkinesia and apraxia, while Parkinson's disease causes bradykinesia gait \cite{cicirelli_human_2022}. It is also found that certain injuries can affect the gait in a particular way \cite{zeng_detecting_2020}. This makes gait a useful and inexpensive biomarker that can be used to diagnose patients, or even to predict the onset of neurological conditions \cite{terrier_gait_2020}. Gait analysis has also been shown to be useful for rehabilitation purposes \cite{mirek_assessment_2016}\cite{prasanth_wearable_2021}. 

Gait analysis can be used in sports for analyzing a player's or athlete's body motion and then, providing feedback. The feedback can be used to improve the player's technique and to avoid injuries during play. In \cite{anand_wearable_2017}, a wrist worn IMU sensor was used for shot detection and classification in swing sports, such as, badminton and tennis. Football shot and pass detection using IMU sensors attached to the shoes worn by players was studied in \cite{stoeve_laboratory_2021}. There is, of course, a growing trend in the tech world towards smartwatch based health and fitness. 

When it comes to HCI, gait analysis has been used for converting human gait into appropriate action in games and Virtual/Augmented Reality (VR/AR). Zhang et al. \cite{zhang_deep_2020} developed triboelectric smart socks to perform gait analysis using deep learning (DL). The system was also demonstrated by using it in a VR health fitness application. 

In biometrics and cybersecurity, gait has been proposed as a method to identify and
authenticate individuals. Gait can be used for easy person identification and re-identification in public spaces such as banks, airports, railway stations, shopping malls, etc. due to its non-invasive nature and without privacy concerns. In \cite{terrier_gait_2020}, a CNN algorithm was trained to recognize individual gait using Ground Reaction Force (GRF) sensors embedded in a treadmill. 

Human Activity Recognition has applications in the areas of elderly care, rehabilitation and personal fitness. In \cite{ashry_lstm-based_2024}, the researchers classified several Activities of Daily Living (ADL), such as, drawing, cycling, work out, etc. using data from an IMU in a smartwatch worn by volunteers. Researchers in \cite{murad_deep_2017} developed a bidirectional LSTM-based deep Recursive Neural Network (RNN) for classifying activities from variable length data generated from IMU datasets. 

Coming to the sensing methods used in gait analysis, the state-of-the-art gait analysis techniques majorly use non-wearable sensors \cite{godinho_systematic_2016}. GAITRite \cite{montero-odasso_quantitative_2009} is a commercial system which is an electronic walkway mat with embedded pressure sensors. The gait data from the pressure sensors is processed to output spatial and temporal parameters, such as, gait velocity, step length, stride length, etc. Three dimensional motion capture systems that use optical systems, sometimes in combination with pressure sensors, are commonly used to create publicly available gait datasets \cite{david_human_2024}\cite{montero-odasso_quantitative_2009}\cite{beauchet_poor_2016}. 

IMUs are the most commonly used sensor in wearable gait analysis systems \cite{prasanth_wearable_2021}. APDM technologies \cite{noauthor_comprehensive_2022} developed IMU modules that can be worn at various locations on the body. Their data acquisition and analysis software called Mobility Lab is able to estimate various gait parameters such as, gait speed, circumduction, dorsiflexion, plantarflexion, stride length, etc. 

Non-wearable systems tend to offer better gait analysis performance compared to wearables \cite{harris_survey_2022}. The greater sophistication in non-wearable systems helps to capture the human gait more effectively, especially using 3D motion capture systems that provide 6DoF tracking data. However, non-wearable systems require expensive setups in clinical settings, which are not scalable and accessible to a large population. Hence, there is a push for the betterment of wearable systems. Current systems that use IMUs are unable to achieve better performance due to the noise and drift in the sensors. This leads to noise and drift in tracking systems that make use of IMUs. Zero velocity update (ZUPT) is a method used to counter this drift issue to a certain extent by re-initializing the integration process during the double stance gait phase \cite{hannink_mobile_2018}\cite{pierleoni_validation_2019}\cite{zhang_accurate_2020}. Optical sensors are typically not used in wearable systems as they need line-of-sight (LoS) to function. Also, optical systems tend to be power intensive and have privacy concerns. Magnetic tracking presents a solution to these issues.
\begin{itemize}
  \item It can work in non line-of-sight (NLos) conditions.
  \item It consumes less power compared to optical systems.
  \item It provides higher update rate due to simpler computation requirements compared to optical systems.
  \item It does not have privacy concerns like optical systems do. 
  \item It does not suffer from drift issues like IMU based tracking systems.
\end{itemize}

In \cite{golestani_human_2020}, signals generated from magnetic induction was used to perform HAR. Multiple transmitter coils (transmitting at different frequencies) could be worn at various locations in the body, which induce voltages in a coil around the torso. This voltage was then fed to a deep RNN model to recognize activities, without actually performing any position tracking. However, the system was mostly simulated assuming ideal conditions. The different shapes and sizes of the coils that would arise due to placing the coils at different locations on the body and across different body shapes was not considered. The shape and size could also vary dynamically as the person moves, which would further complicate the system. The researchers then developed a 3D tracking system using the same concept of magnetic induction in \cite{golestani_wearable_2021}. The tracking was performed using a data driven approach by feeding the induced voltages into a Machine Learning (ML) model and training it using supervised learning. However, the issue of symmetry in magnetic tracking that gives rise to multiple possible solutions was taken care of by constraining the tracking domain. Also, the system does not perform orientation tracking, which could hold important gait information.

We present here a gait analysis system that is based on our previously developed magnetic tracking system that performs 6DoF tracking \cite{singh_magnetic_2020}\cite{singh_inside-out_2021}. To the best of our knowledge, this is the first work to prove the feasibility of using 6DoF wearable magnetic tracking system for gait analysis. The system consists of receiver (Rx) modules whose position and orientation is tracked with respect to a transmitter (Tx) module. The Tx module sets up a magnetic field which is sensed by tri-axis coils on the Rx module. All the modules also consist of IMUs and magnetometers to find their orientation. The system is portable, power efficient and minimally intrusive. In the present work, we make use of this magnetic tracking system for the application of HAR. Twelve participants were made to perform 4 activities - walking, walking with weight, jogging and marching on the spot (henceforth, referred to as just marching) - while wearing the magnetic tracking system. The classification was performed using CNN and LSTM classifiers. Only DL classifiers have been considered as they have shown superior performance compared to conventional ML classifiers \cite{tunca_deep_2020}\cite{shao_multi-modal_2022}\cite{murad_deep_2017}\cite{stoeve_laboratory_2021}\cite{tabrizi_comparative_2020}\cite{burdack_systematic_2020}. Surveys of all the gait analysis methods is taken up in \cite{harris_survey_2022}\cite{khera_role_2020}\cite{prasanth_wearable_2021}\cite{lara_survey_2013}. For comparison with the commonly used wearable sensors - IMU and magnetometer - we repeated the experiment by placing an IMU and a magnetometer on each foot of the participants. The magnetic tracking system shows a significantly higher classification accuracy as compared to the IMU based system. This is owed to the fact that IMUs cannot strictly perform 6DoF tracking, thus, providing incomplete gait information. 

The paper is organized as follows. Section II presents the methodology. Here, a brief description of the magnetic tracking system is mentioned for convenience. More details about the tracking system can be found in \cite{singh_magnetic_2020}\cite{singh_inside-out_2021}. The data collection and preprocessing methods are described. Then, a description of the ML methods used is also provided. The activity recognition results are presented in Section III. Following that, discussion and conclusions are presented in Section IV and Section V.

\section{Methodology} 
\label{methodology}
\subsection{Magnetic tracking system}
\label{mag_track_sys}
\begin{figure}[t]
\centering
\includegraphics[width=10.1cm]{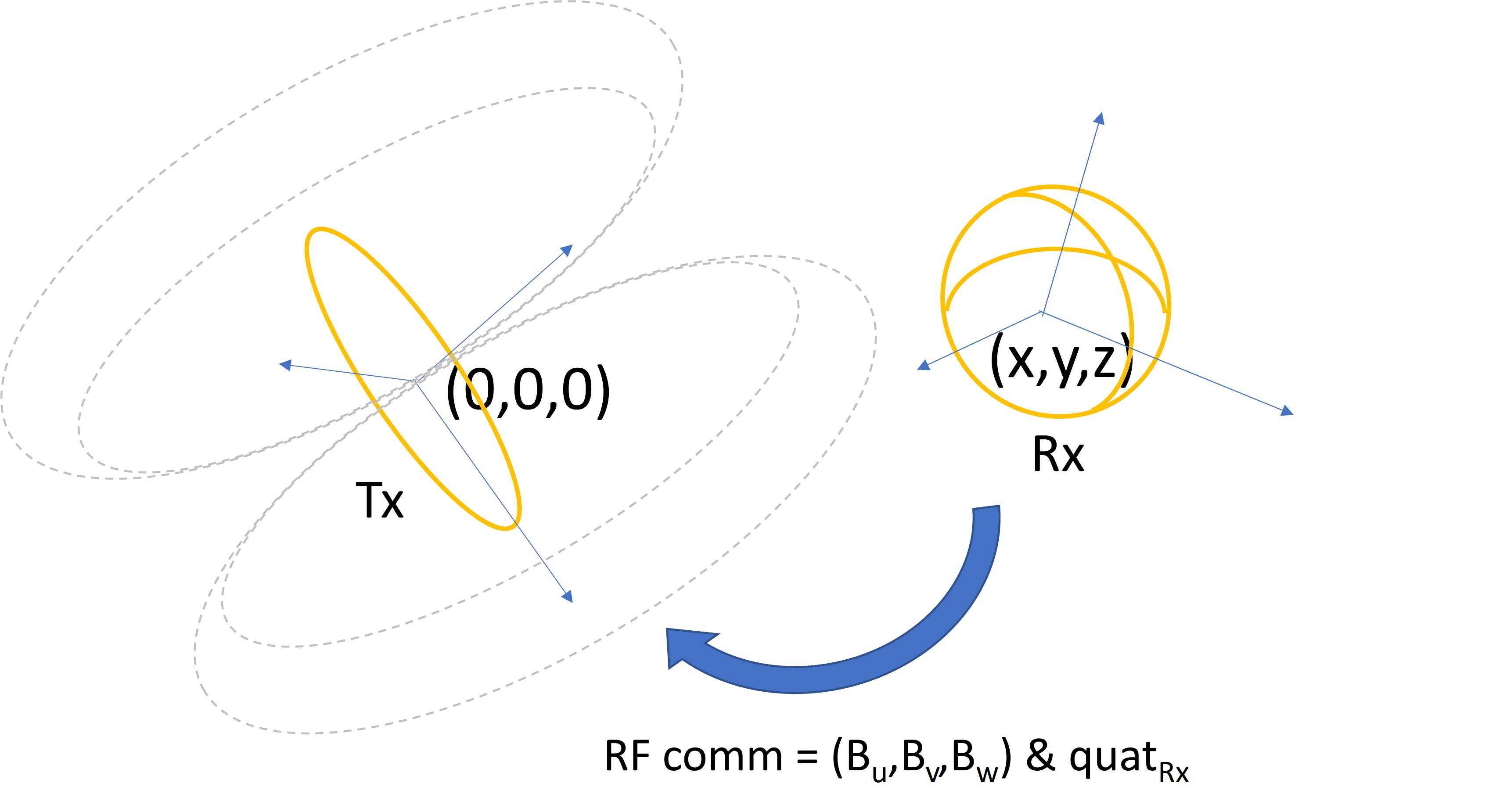}
\caption{The tri-axis Rx coil senses the magnetic field generated by the Tx coil centered at the origin. The Rx then communicates the magnetic field vector to the Tx, along with its orientation. The Rx module's position and orientation are estimated with respect to the Tx module \cite{shankar_calibration_2024}.}
\label{magnetic_tracking_sys}
\end{figure}

We present here a brief description of our magnetic tracking system \cite{singh_magnetic_2020}. An Rx module is tracked with respect to a Tx module. The Rx module mainly consists of a tri-axis coil, IMU and magnetometer. The Tx module mainly consists of an AC magnetic field generator coil, IMU and magnetometer. IMU and magnetometer present on each module is used to find the orientation of that module with respect to earth. The position finding problem can be visualized in Fig \ref{magnetic_tracking_sys}. 

The Tx module sets up a magnetic field as per the point source model of magnetic field in the Tx reference frame \cite{singh_magnetic_2020}:
\begin{equation}
B_x=  \frac{3Mxz}{4\pi r^5 } \label{Bx}
\end{equation}
\begin{equation}
B_y=  \frac{3Myz}{4\pi r^5 } \label{By}
\end{equation}
\begin{equation}
B_z=  \frac{2M(2z^2-x^2-y^2 )}{4\pi r^5}\label{Bz}
\end{equation}
where \((x, y, z)\) are the position coordinates to be found, \(r= \sqrt{x^2+y^2+z^2}\), M = magnetic moment. Having known (B\textsubscript{x}, B\textsubscript{y}, B\textsubscript{z}), the position \((x, y, z)\) can be calculated using a closed form solution.

\begin{figure} [h]
    \centering
    \includegraphics[width=8.8cm]{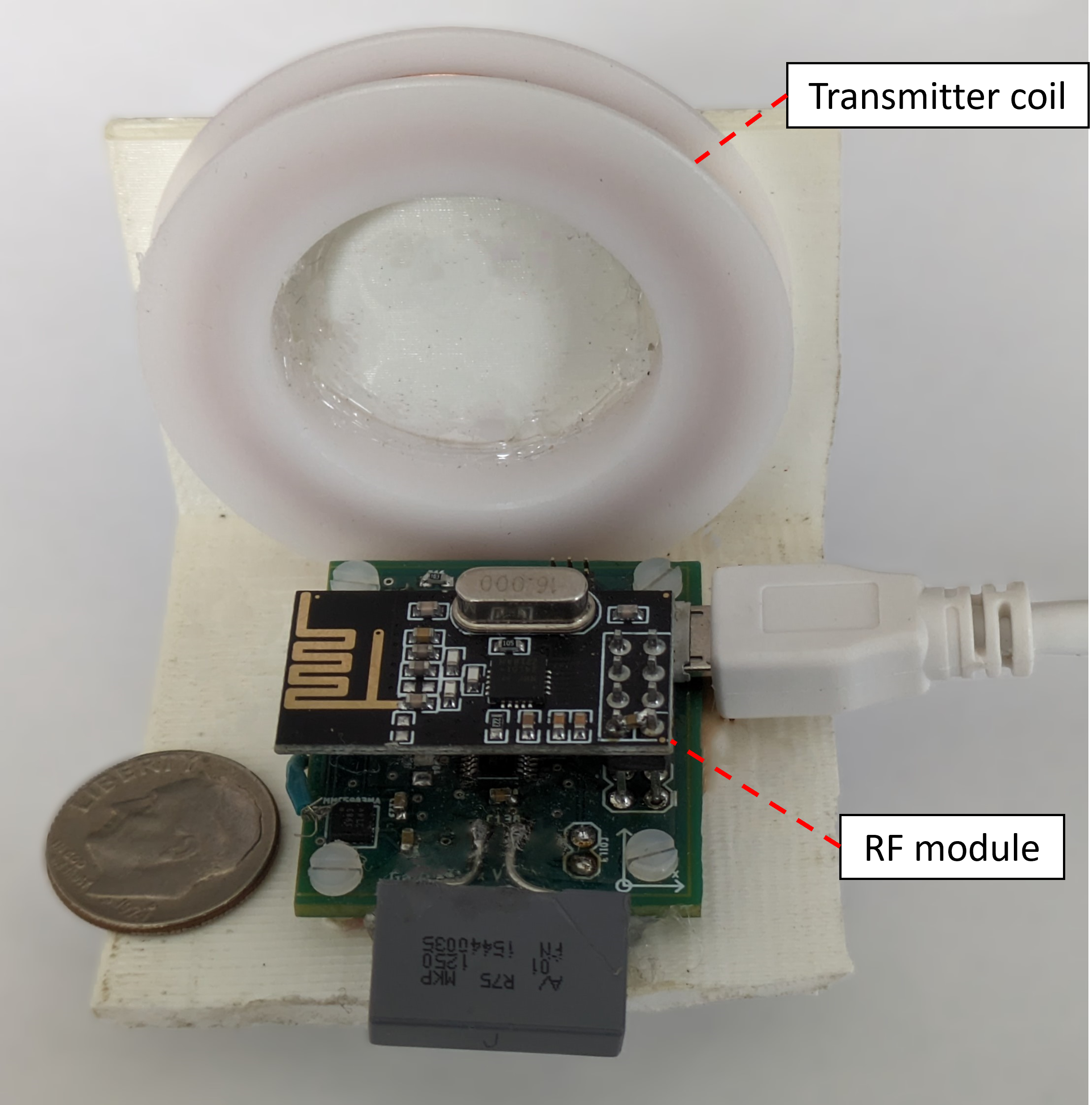}
    \caption{The Tx module mainly consists of the transmitter coil to generate the magnetic field. It receives the data from the Rx module using the RF communication module. It transfers the position and orientation through the USB.}
    \label{tx_module}
\end{figure}

\begin{figure} [h]
    \centering
    \includegraphics[width=8.8cm]{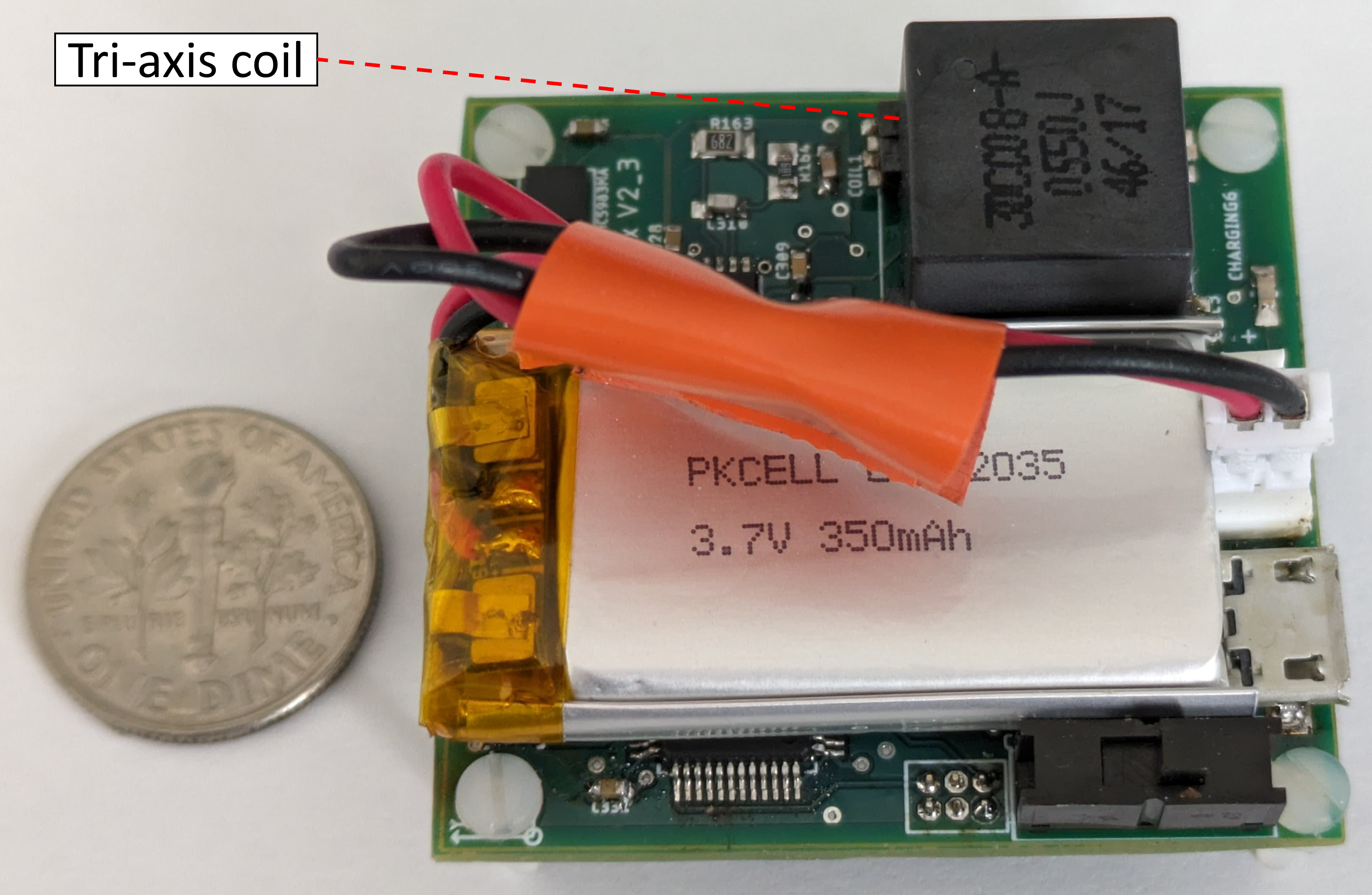}
    \caption{The Rx module mainly consists of a tri-axis coil to sense the magnetic field generated by Tx.}
    \label{rx_module}
\end{figure}

Referring to Fig \ref{magnetic_tracking_sys}, the Rx module senses the amplitude of the magnetic field using the tri-axis coil, to get a magnetic field vector in the Rx reference frame. This magnetic field vector, along with the Rx orientation is communicated to the Tx module. The Tx module finds the relative orientation and rotates the magnetic field vector from the Rx reference frame to the Tx reference frame. Thus, (B\textsubscript{x}, B\textsubscript{y}, B\textsubscript{z}) is obtained, which is then used in the point source model to back calculate the position. This way, the tracking system outputs the position and orientation (as quaternions) of the Rx module with respect to the Tx module. The Tx and Rx modules can be seen in Fig \ref{tx_module} and Fig \ref{rx_module}. For more details about the tracking system, the reader is referred to \cite{singh_magnetic_2020}\cite{singh_inside-out_2021}.


\subsection{Data collection}
\label{data_collection}
\begin{figure} [h]
    \centering
    \includegraphics[width=5.6cm]{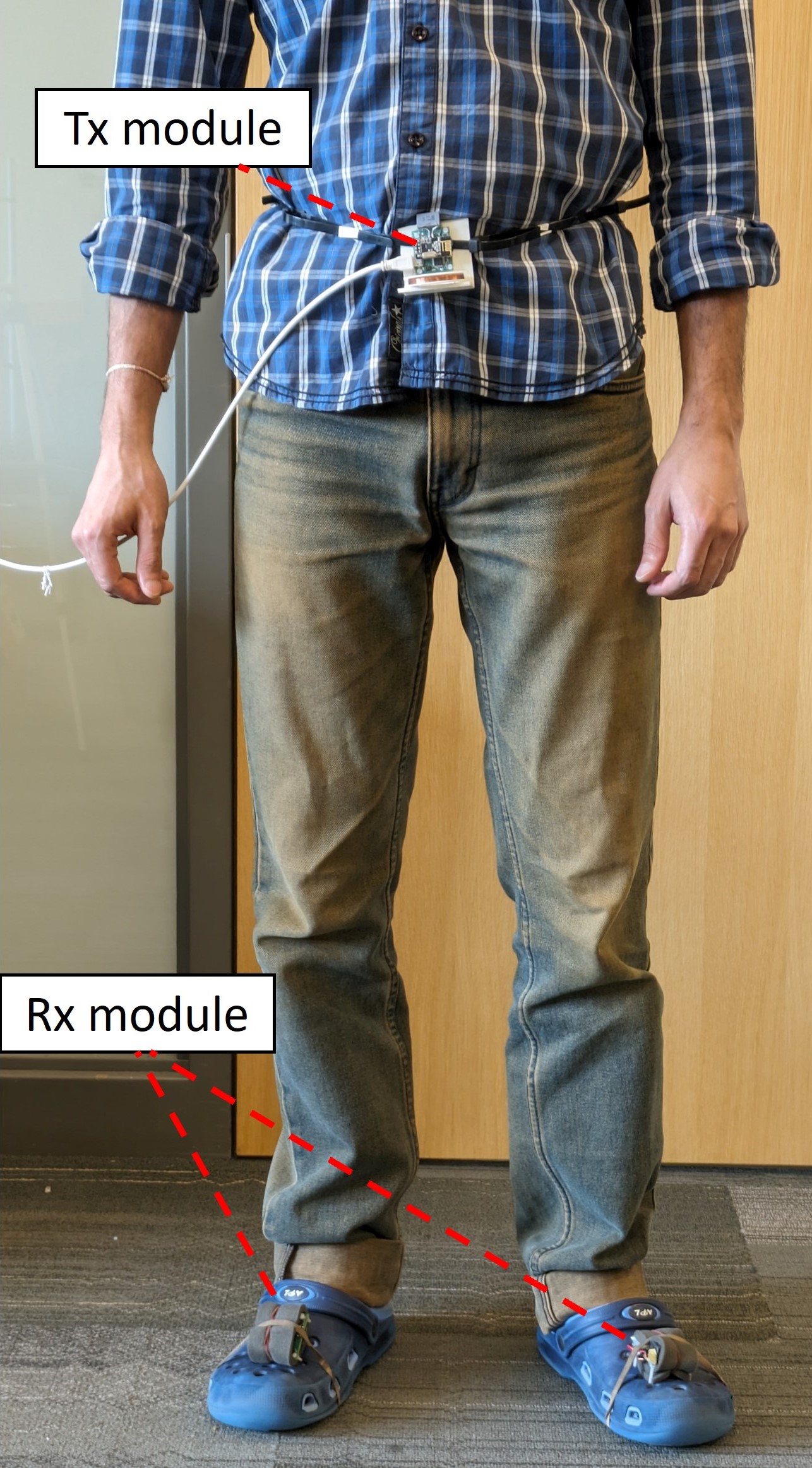}
    \caption{The setup that was used for collecting gait. The tracking of the Rx modules on the feet is performed with respect to the Tx module that is placed on the waist.}
    \label{gait_setup}
\end{figure}

The gait data of 12 participants was collected in two phases. In the first phase, the gait data was collected using the magnetic tracking system. The Tx module was placed near the lower part of the abdomen, secured with a belt around the waist. The position and orientation data of the Rx modules is output using the USB cable attached to Tx, and logged onto a laptop computer using serial communication. An Rx module was placed on the dorsum of each foot - Rx1 on left foot and Rx2 on right foot. The setup is shown in Fig \ref{gait_setup}. The data from Rx1 and Rx2 are communicated asynchronously to the Tx using an RF communication module at roughly 300 samples/sec. The gait data of 12 participants - 9 males and 3 females - was collected. Each participant performed 4 activities: 
\begin{itemize}
  \item Walking (W) - participants were free to walk in a straight line at a self-chosen speed for 10 seconds at a time.
  \item Walking with weight (WW) - participants walked similar to W but carrying a backpack weighing 8 kgs.
  \item Jogging (J) - participants performed a very low intensity jog for 10 seconds at a time.
  \item Marching on spot (M) - participants marched on the spot for 10 seconds at a time. 
\end{itemize}
Each participant was asked to perform each activity for 10 seconds at a time for a total of 1 minute duration for each activity.  

We compare the magnetic tracking system with an IMU + magnetometer based system, which is currently the most common sensing method using wearables. So, in the second phase of data collection, IMUs and magnetometers present on the same system were used. Here, the gait data of a new set of 12 participants - 9 male and 3 female - was collected. The exact data collection protocol as that for the first phase was followed. All the data was collected in the same spot in an open parking lot, away from metallic objects. This is because magnetometers and hence, the magnetic tracking system are affected by metals distorting magnetic fields.

\subsection{Preprocessing}
\label{preprocessing}
\begin{figure*} [h]
    \centering
    \includegraphics[width=19.5cm]{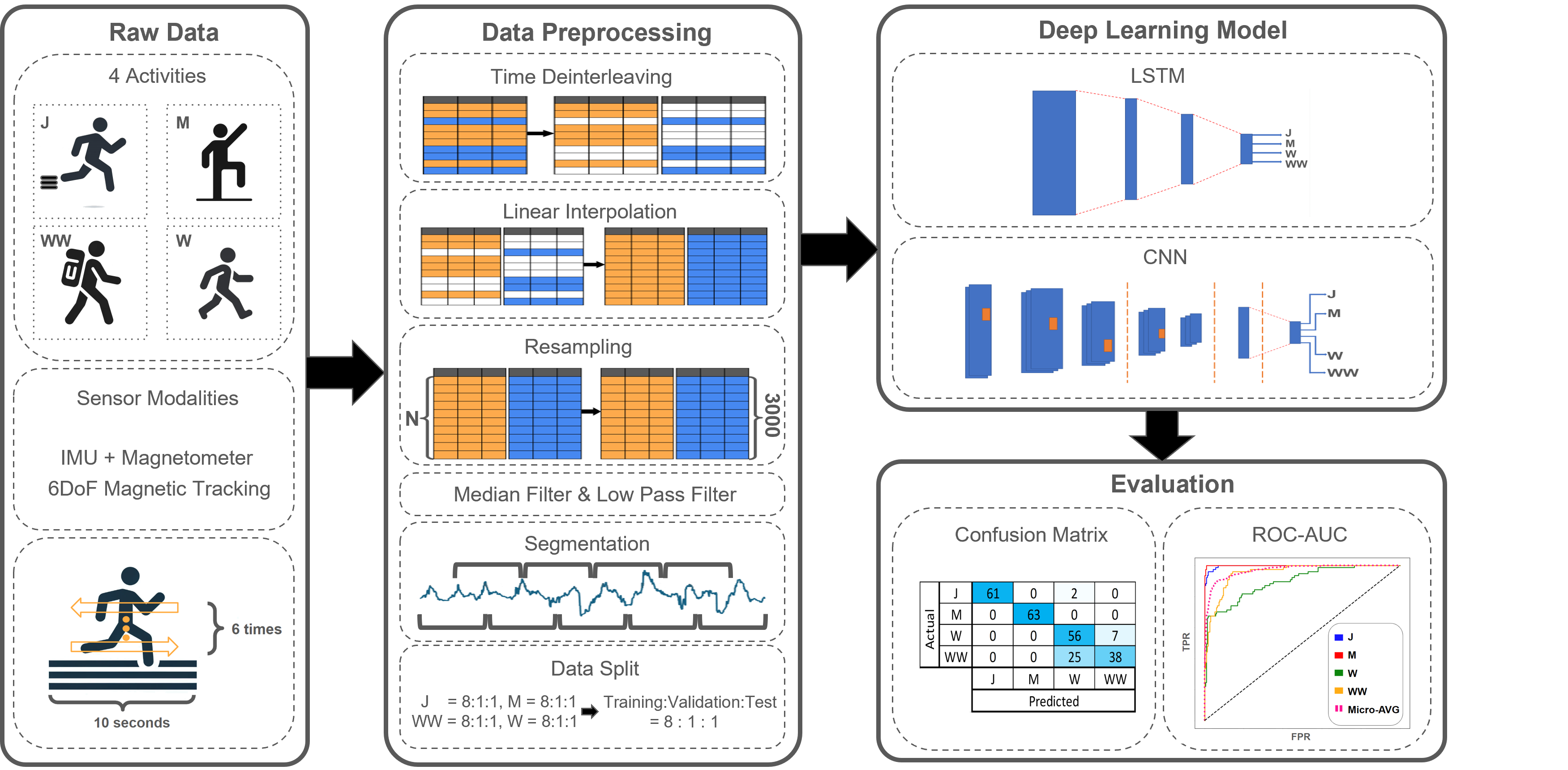}
    \caption{The overall block diagram for the HAR system including the data collection, preprocessing steps, DL classifiers and the evaluation metrics.}
    \label{overall_block}
\end{figure*}

The overall block diagram for the data pipeline is show in Fig \ref{overall_block}. The data from Rx1 and Rx2 are received by Tx asynchronously at roughly 300 samples/sec. Data packets received by Tx from Rx1 and Rx2 had ID numbers 1 and 2. These packets are randomly interleaved as shown in Fig \ref{overall_block}. The packets are de-interleaved and the missing data points (both position and orientation) for Rx1 and Rx2 are linearly interpolated. This would now give us equal number of data points for Rx1 and Rx2. A median filter of window size 5 was applied at this stage to remove any outlier spikes in the data caused mostly by communication issues or device malfunctions. The data points are then resampled to give us 3000 points in every 10 second data file, which amounts to precisely 300 samples/sec update rate. 

\begin{figure} [h!]
    \begin{subfigure}[b]{0.3\textwidth}
        \includegraphics[width=9.1cm]{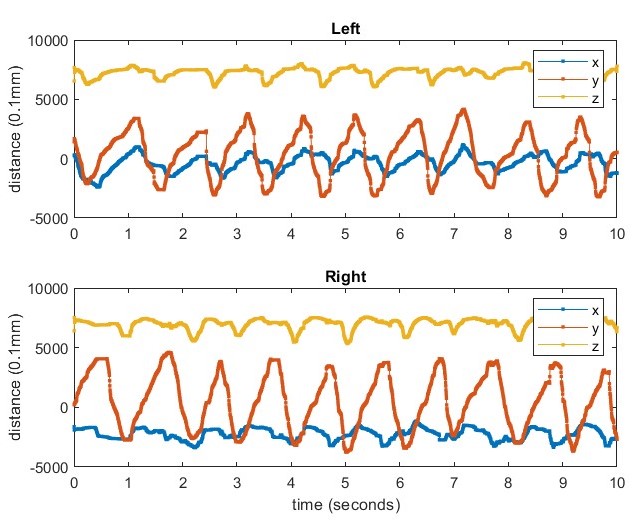}
        \caption{Position data.}
    \end{subfigure}
    \hfill
    \begin{subfigure}[b]{0.3\textwidth}
        \includegraphics[width=9.1cm]{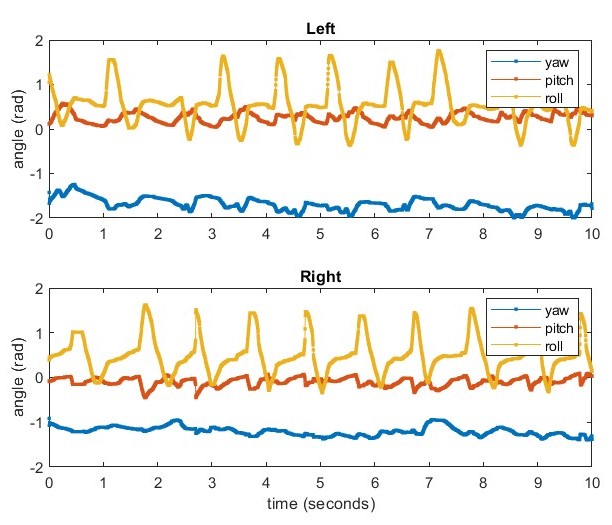}
        \caption{Orientation data.}
    \end{subfigure}
    \caption{Example of a 10s (a) position and (b) orientation tracking data of the left foot (top) and right foot (bottom) during walking.}
    \label{w_example}
\end{figure}

Note that, the orientation is represented as a quaternion so far. We convert it to Euler angles as they are a better intuitive representation for orientation, making it easier to learn for ML. This step is not applicable to the data from the IMU + magnetometer system. A low pass elliptic filter with a cut off frequency of 15 Hz was applied on the time series position and orientation data. An example of the position data of the left and right feet collected for 10s of the walking task is shown in Fig \ref{w_example}.

The data was segmented into smaller time windows. We tried time windows of length 2 seconds (600 time samples) and 1.67 seconds (500 time samples) in line with the conclusion in \cite{banos_window_2014}. We applied a sliding time window with a slide amount of 1 second, giving us 9 data windows for every 10 second data window. This would help create more data samples. Also, the ML model is to be trained to be time shift invariant. Finally, min-max normalization was applied on each window to have the data scale between 0 and 1 for efficient ML training. All the preprocessing was performed using MATLAB.

After preprocessing, for each activity, we have about 648 data samples (12 people \(\times\) 6 10s data windows \(\times\) 9 2s data windows) for 2 second time windows and 792 data windows (12 people \(\times\) 6 10s data samples \(\times\) 11 1.67s data windows) for 1.67 second time windows. For the magnetic tracking data, the size of each 2s data sample is 600\(\times\)12 and that of the 1.67s data sample is 500\(\times\)12. For the IMU + magnetometer data, the size of each 2s data sample is 600\(\times\)18 and that of the 1.67s data sample is 500\(\times\)18. The feature sizes can be understood using Table \ref{tab_mag_track} and Table \ref{tab_IMU_magno}. 

\begin{table}[]
\renewcommand{\arraystretch}{1.5}
\caption{12 features in magnetic tracking gait data.}
\label{tab_mag_track}
\begin{tabular}{|llllll|llllll|}
\hline
\multicolumn{6}{|c|}{\textbf{Left}}                                                                                                               & \multicolumn{6}{c|}{\textbf{Right}}                                                                                                              \\ \hline
\multicolumn{3}{|c|}{Position}                                            & \multicolumn{3}{c|}{Orientation}                             & \multicolumn{3}{c|}{Position}                                            & \multicolumn{3}{c|}{Orientation}                             \\ \hline
\multicolumn{1}{|l|}{x} & \multicolumn{1}{l|}{y} & \multicolumn{1}{l|}{z} & \multicolumn{1}{l|}{yaw} & \multicolumn{1}{l|}{pitch} & roll & \multicolumn{1}{l|}{x} & \multicolumn{1}{l|}{y} & \multicolumn{1}{l|}{z} & \multicolumn{1}{l|}{yaw} & \multicolumn{1}{l|}{pitch} & roll \\ \hline
\end{tabular}
\end{table}

\begin{table}[]
\caption{18 features in IMU + magnetometer gait data. Gyro = gyroscope; Accel = accelerometer; Magno = magnetometer.}
\label{tab_IMU_magno}
\resizebox{8.9cm}{!} {
\renewcommand{\arraystretch}{2}
\begin{tabular}{|lllllllll|lllllllll|}
\hline
\multicolumn{9}{|c|}{\textbf{Left}}                                                                                                                                                                                 & \multicolumn{9}{c|}{\textbf{Right}}                                                                                                                                                                                \\ \hline
\multicolumn{3}{|c|}{Gyro}                                                & \multicolumn{3}{c|}{Accel}                                               & \multicolumn{3}{c|}{Magno}                          & \multicolumn{3}{c|}{Gyro}                                                & \multicolumn{3}{c|}{Accel}                                               & \multicolumn{3}{c|}{Magno}                          \\ \hline
\multicolumn{1}{|l|}{x} & \multicolumn{1}{l|}{y} & \multicolumn{1}{l|}{z} & \multicolumn{1}{l|}{x} & \multicolumn{1}{l|}{y} & \multicolumn{1}{l|}{z} & \multicolumn{1}{l|}{x} & \multicolumn{1}{l|}{y} & z & \multicolumn{1}{l|}{x} & \multicolumn{1}{l|}{y} & \multicolumn{1}{l|}{z} & \multicolumn{1}{l|}{x} & \multicolumn{1}{l|}{y} & \multicolumn{1}{l|}{z} & \multicolumn{1}{l|}{x} & \multicolumn{1}{l|}{y} & z \\ \hline
\end{tabular}
}
\end{table}

All data windows are labelled according to the activity - J (0), M (1), W (2), WW (3). All the time windows for all the activities are combined and shuffled. The data set is split into training, validation and test sets in the ratio 80:10:10.

\subsection{Activity recognition using Deep Learning}
\label{act_recognition}
We compared the performance of 2 commonly used DL classifiers - CNN and LSTM. DL classifiers take the input gait data with minimal preprocessing. They are able to extract the discriminative features automatically without requiring the need for feature engineering that is needed for conventional ML models like Support Vector Machines (SVM) and Multi-Layer Perceptron (MLP). Feature engineering requires domain expertise which may not be feasible to avail for all applications. Even with feature engineering, previous gait analysis research has shown that DL models, given enough data, tend to outperform conventional ML models \cite{stoeve_laboratory_2021}\cite{tabrizi_comparative_2020}\cite{burdack_systematic_2020}.

The Convolution Neural Network (CNN) model was developed taking inspiration from the way human brain processes visual information \cite{tabrizi_comparative_2020}. Kernels of a certain size are convolved with the input data sample to generate convolution feature outputs. The size of this output is reduced by discarding some information, using some method such as the commonly used MaxPooling method. The convolution and maxpooling operations are repeated several times back-to-back. The final output is then input to a fully connected (FC) layer, which outputs the data labels using a softmax layer, for classification \cite{ordonez_deep_2016}. Dropout layers can be added to avoid overfitting. 

Typical ML models do not have a memory element to process sequential data. So, the relationship between semantically, temporally or spatially close data points are not captured. To that end, Recurrent Neural Networks (RNN) was developed where the output of a single cell was dependent on both the current input and an internal cell state, which gets updated at every input sample point \cite{ordonez_deep_2016}. Training a simple RNN cell has the issue of exploding and vanishing gradients \cite{murad_deep_2017}. To resolve that, Long Short Term Memory (LSTM) model was developed as an extension of RNN. LSTM resolves the issue with RNN using gating cells that control the flow of information. LSTM cells sequentially process the input data, one sample at a time, while updating the internal state. At the end of the data sequence, the output of the LSTM is input into a FC layer, which finally outputs the data labels using a softmax layer, for classification. 

\begin{figure} [h]
    \includegraphics[width=8.8cm]{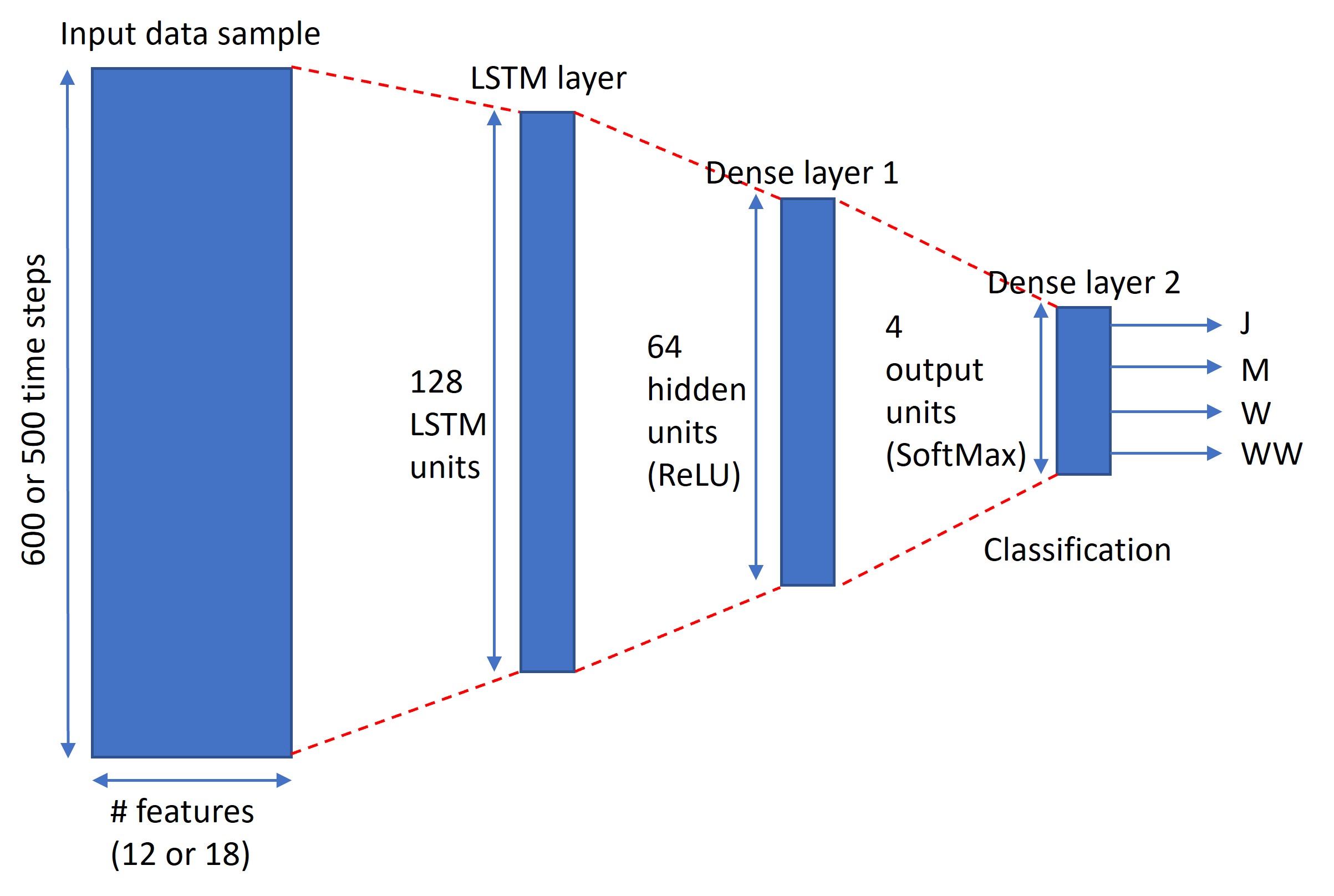}
    \caption{Structure of the LSTM model.}
    \label{lstm_structure}
\end{figure}

\begin{figure} [h]
    \includegraphics[width=8.8cm]{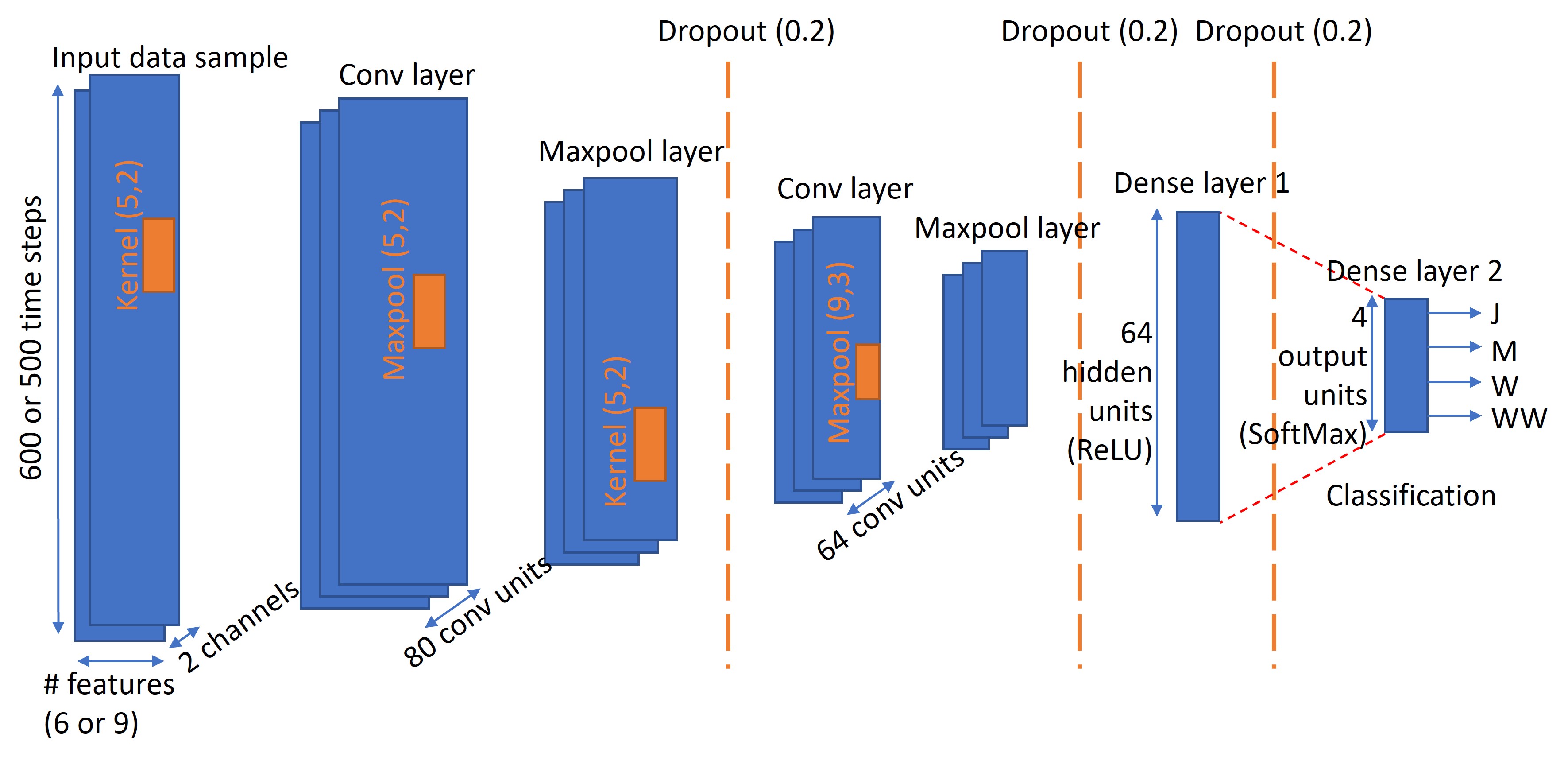}
    \caption{Structure of the CNN model.}
    \label{cnn_structure}
\end{figure}

The models were trained on Python 3.9 using the Tensorflow library. The optimizer 'Adam' was used during training. Models were evaluated using the metric of accuracy. The training and validation data sets were used for the hyperparameter tuning. Hyperparameter tuning was performed separately for magnetic tracking data and IMU + magnetometer data. For the CNN model, the data from left and right foot were input as 2 different channels. We performed parametric sweeps for the number of convolution units, epochs, batch size and size of the dense units. We chose the CNN model that gave the best validation accuracy. The hyperparameters of the chosen CNN model is shown in Fig \ref{cnn_structure}. For the LSTM model, we performed parametric sweeps for the number of LSTM units, batch size, epochs and size of the dense units. We chose the LSTM model with the best validation accuracy. The hyperparameters of the chosen LSTM model is shown in Fig \ref{lstm_structure}.

\section{Results}
\label{results}
The entire data set was split as training, validation and testing data sets in the ratio 8:1:1. Only the training and validation data sets were used for the hyperparameter tuning. The testing data set, which was not seen by the ML models, was used for the evaluation. Once the hyperparameters were chosen, the models were trained again with the combined training + validation data set and evaluated on the test data set. This was repeated 8 times for each ML model by shuffling the same training + validation data set with a random seed at each iteration. This helps us evaluate the robustness of the ML models and the gait data. 

The mean accuracy and the standard deviation of the 8 runs is shown in Table \ref{tab_accuracy}. The best performance of 92\% overall mean accuracy is shown by the LSTM model using the magnetic tracking data while segmenting the input using 1.67s moving window. This performance is closely followed by the 2s window segmentation. In general, it can be seen that the 1.67s window and 2s window segmentation have similar performance. This is in line with the conclusion in \cite{banos_window_2014}. We also notice that LSTM has a better overall performance compared to CNN. This could be attributed to the fact that LSTM is better able to learn relationships between even temporally far time samples, in comparison to CNN. Most importantly, the magnetic tracking data consistently performs better than the IMU + magnetometer data. There is an average performance improvement of about 4\%. This confirms our hypothesis that the gait information content is higher in 6DoF tracking data as compared to IMU + magnetometer data. 

\begin{table}[]
\caption{The overall mean \% accuracy\(\pm\)standard deviation of LSTM and CNN classifiers for 2s and 1.67s time windows.}
\label{tab_accuracy}
\renewcommand{\arraystretch}{2}
\begin{tabular}{|l|ll|ll|}
\hline
\multirow{2}{*}{} & \multicolumn{2}{l|}{\textbf{Magnetic tracking}}         & \multicolumn{2}{l|}{\textbf{IMU + magnetometer}}           \\ \cline{2-5} 
                  & \multicolumn{1}{l|}{2s window}        & 1.67s window    & \multicolumn{1}{l|}{2s window}        & 1.67s window     \\ \hline
\textbf{LSTM}     & \multicolumn{1}{l|}{90.72\(\pm\)2.38} & \cellcolor{yellow}92\(\pm\)1.19   & \multicolumn{1}{l|}{86.26\(\pm\)3.4}  & 86.69\(\pm\)3.12 \\ \hline
\textbf{CNN}      & \multicolumn{1}{l|}{86.16\(\pm\)1.68} & 87.4\(\pm\)2.27 & \multicolumn{1}{l|}{85.11\(\pm\)3.65} & 83.15\(\pm\)5.17 \\ \hline
\end{tabular}
\end{table}

\begin{figure*}[t]
\begin{tabular}{lcccc}
 & \multicolumn{2}{c}{\textbf{Magnetic tracking}} & \multicolumn{2}{c}{\textbf{IMU + magnetometer}} \\
\textbf{LSTM} &
\includegraphics[width=.21\linewidth,valign=m]{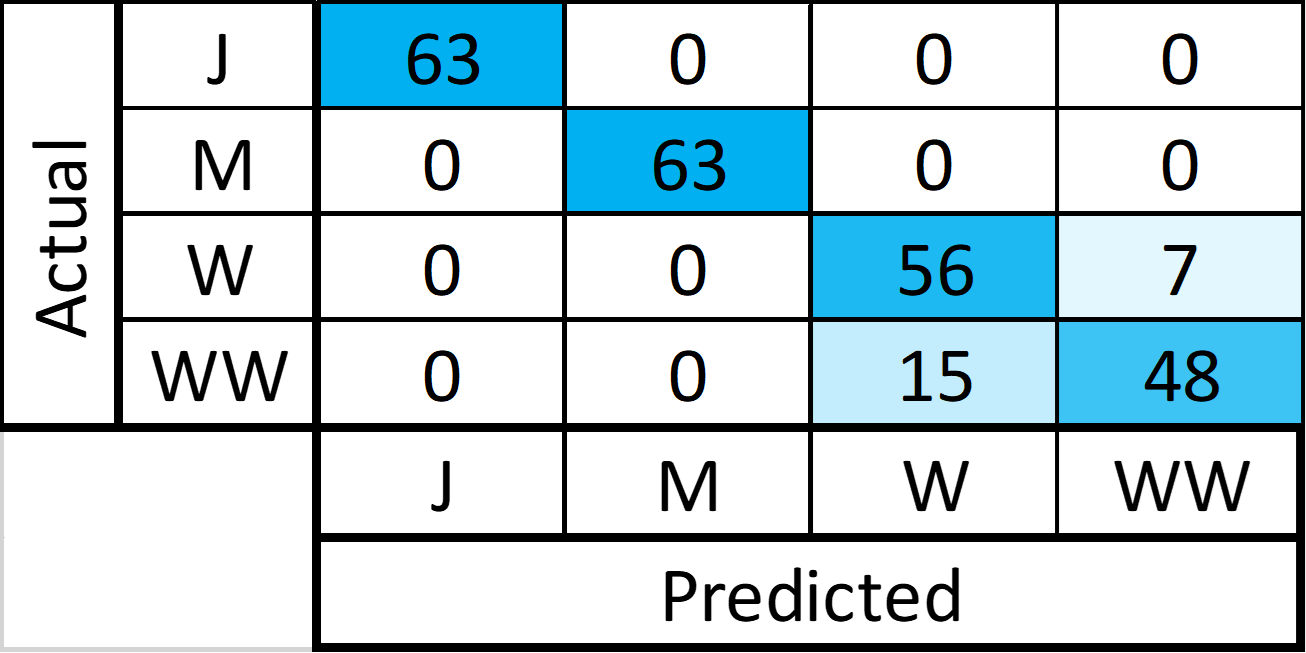} & \includegraphics[width=.21\linewidth,valign=m]{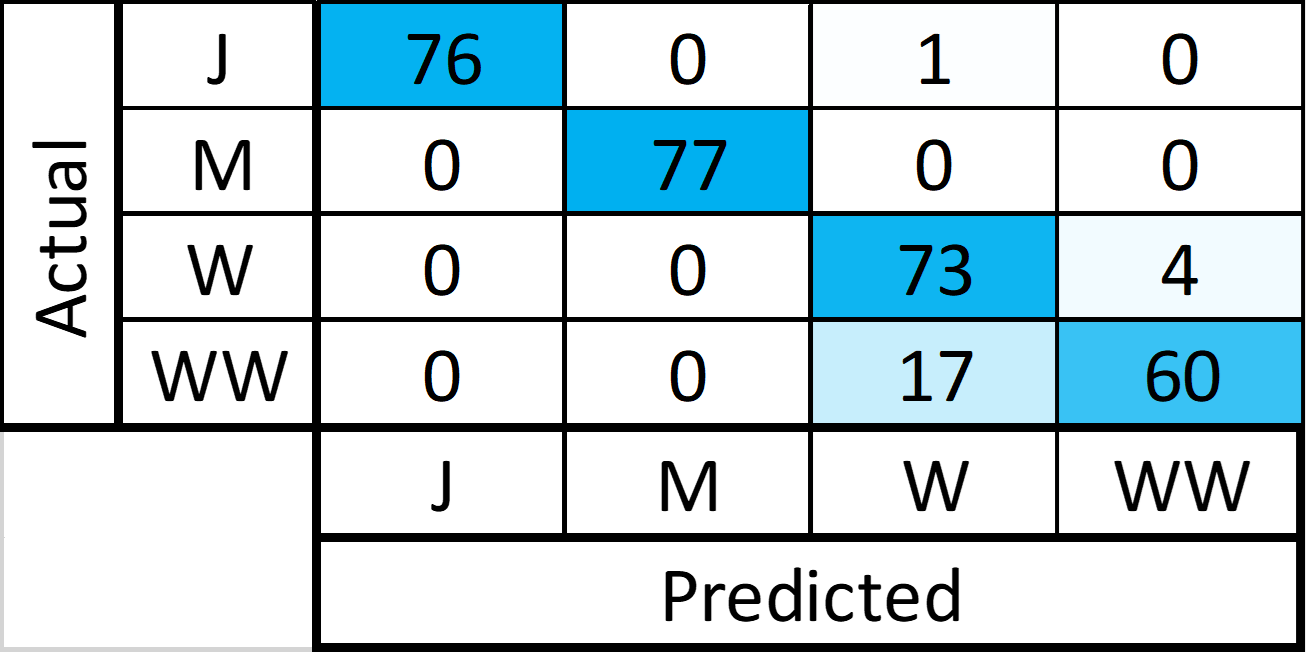} & \includegraphics[width=.21\linewidth,valign=m]{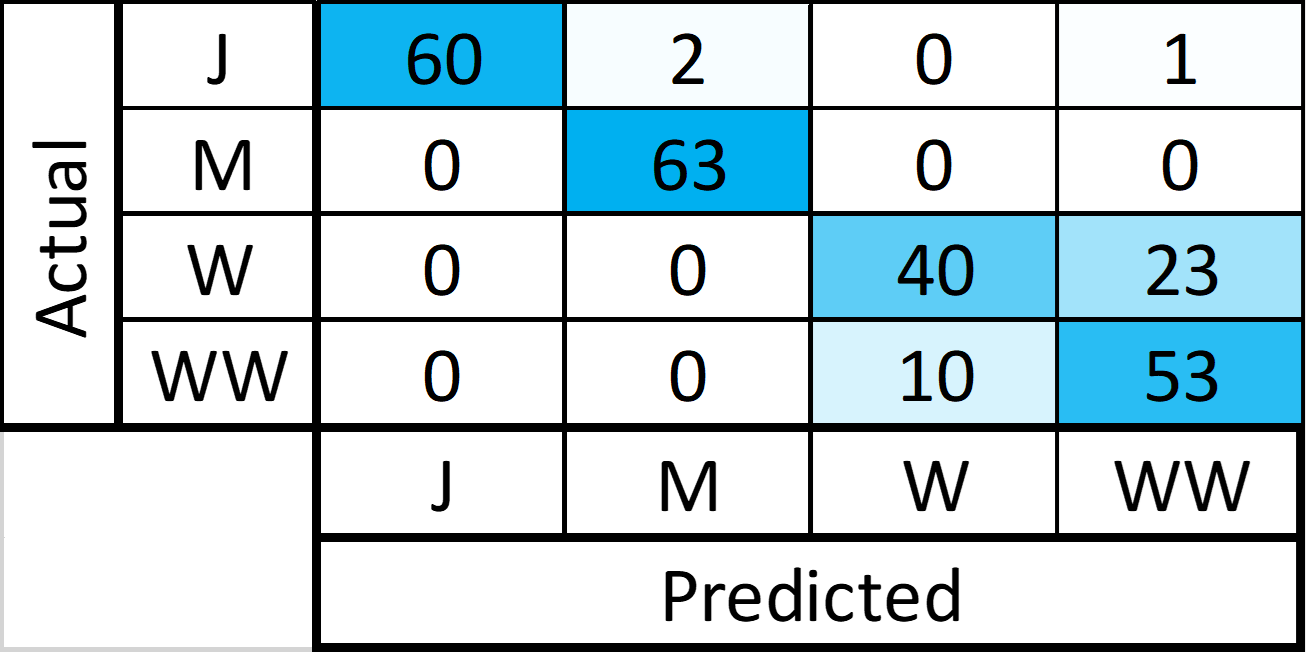} & \includegraphics[width=.21\linewidth,valign=m]{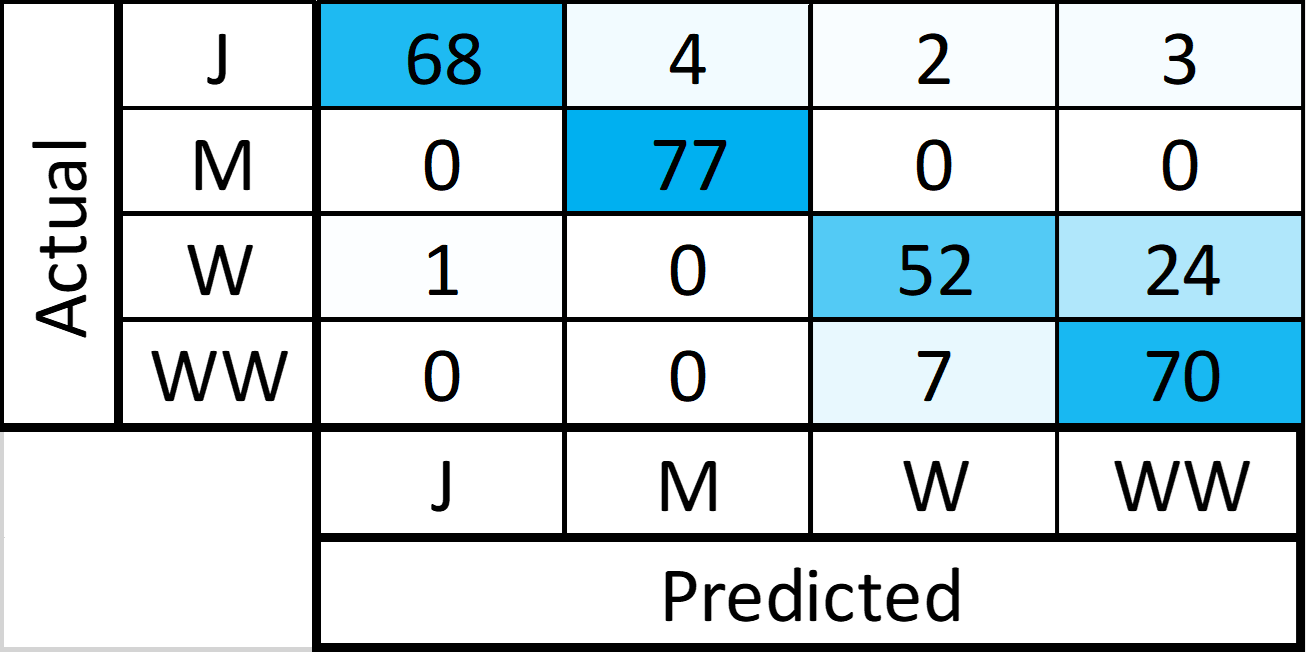}\\
 & (a) & (b) & (c) & (d) \\
\textbf{CNN} &
\includegraphics[width=.21\linewidth,valign=m]{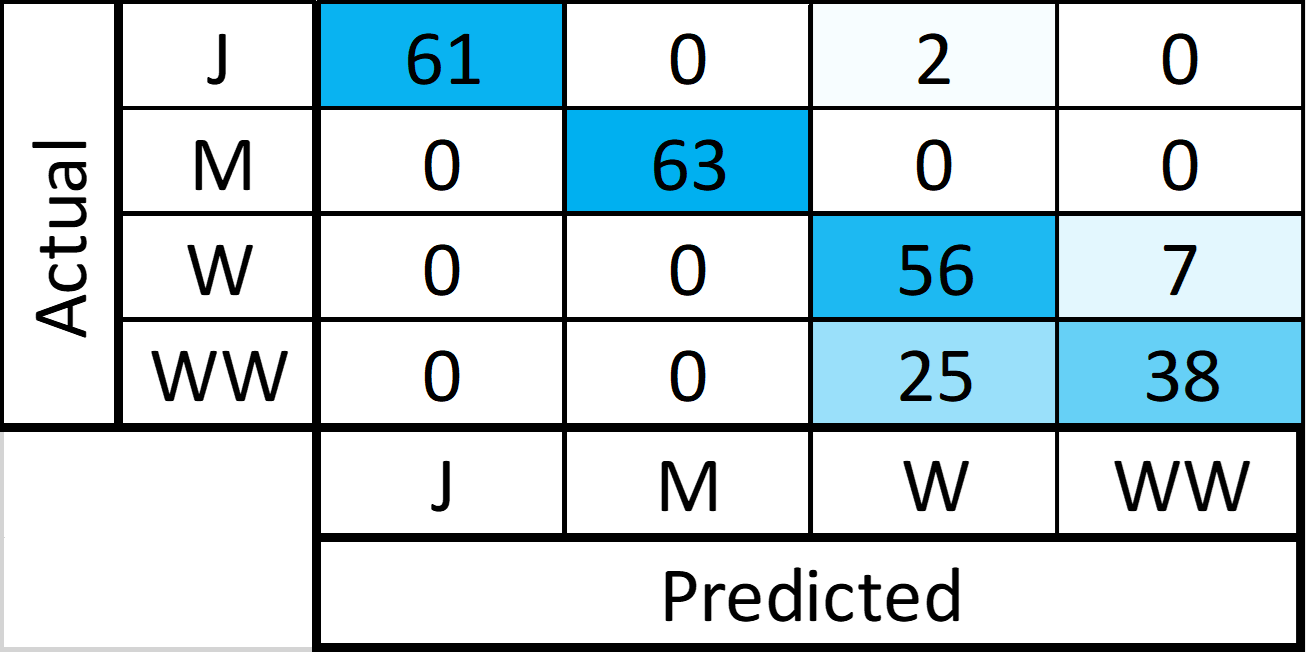} & \includegraphics[width=.21\linewidth,valign=m]{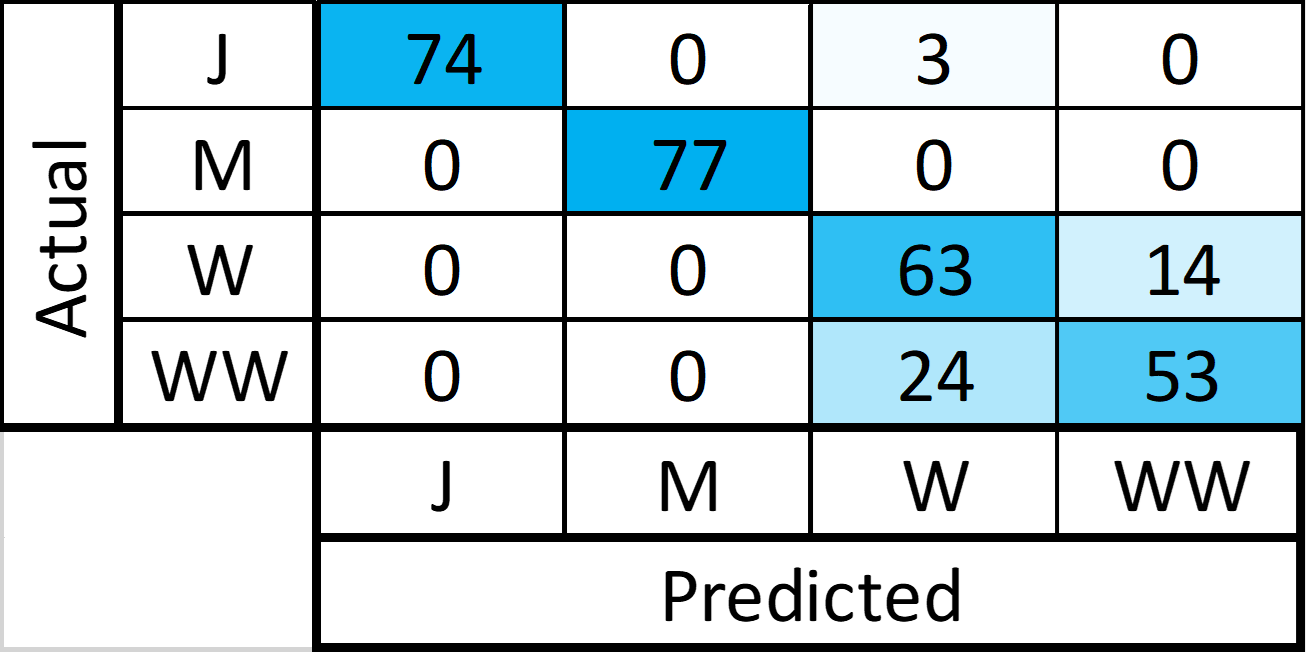} & \includegraphics[width=.21\linewidth,valign=m]{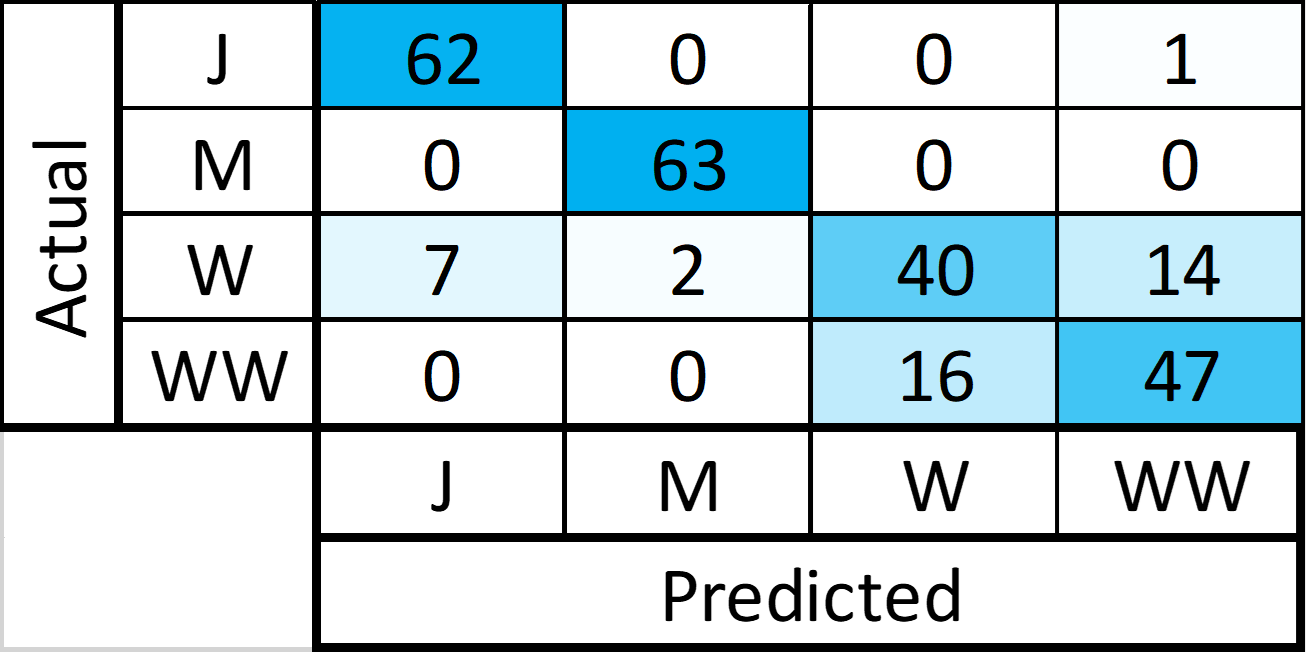} & \includegraphics[width=.21\linewidth,valign=m]{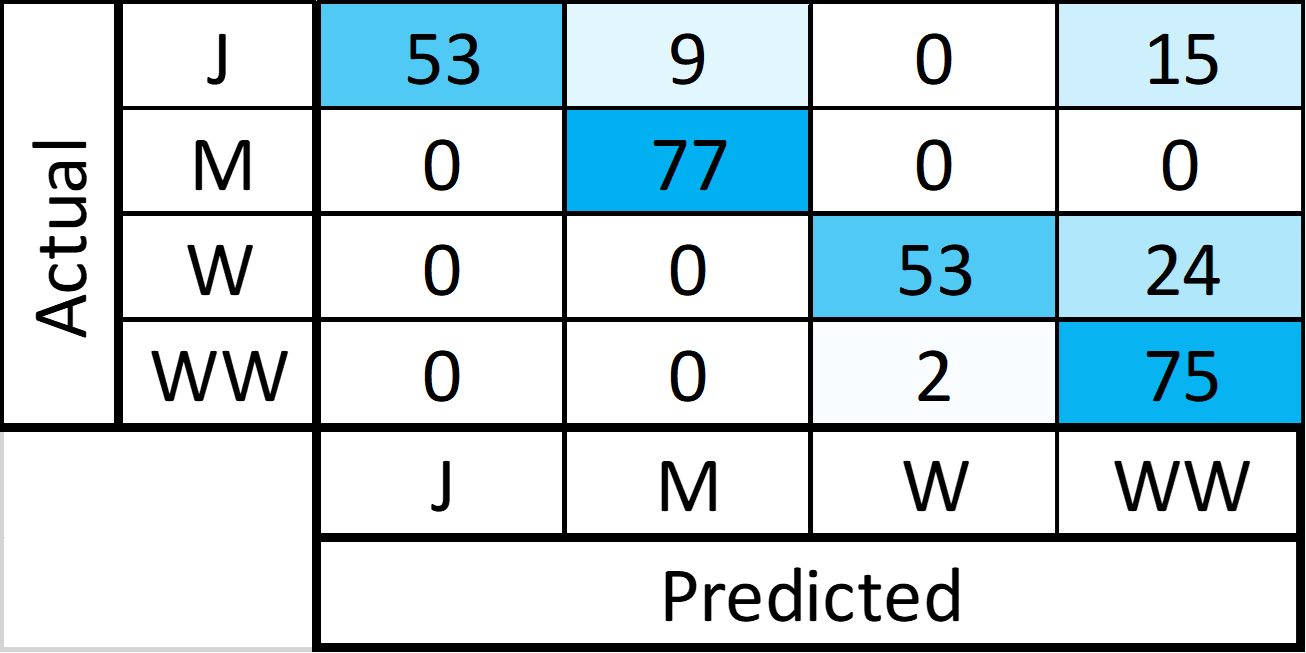}\\
 & (e) & (f) & (g) & (h) \\
\end{tabular}
\caption{Confusion matrix for the classification run closest to the mean accuracy - (a) LSTM, 2s, magnetic tracking, (b) LSTM, 1.67s, magnetic tracking, (c) LSTM, 2s, IMU + magnetometer, (d) LSTM, 1.67s, IMU + magnetometer, (e) CNN, 2s, magnetic tracking, (f) CNN, 1.67s, magnetic tracking, (g) CNN, 2s, IMU + magnetometer, (h) CNN, 1.67s, IMU + magnetometer.}
\label{conf_mat}
\end{figure*}

\begin{figure} [h!]
    \begin{subfigure}[b]{0.3\textwidth}
        \includegraphics[width=8.8cm]{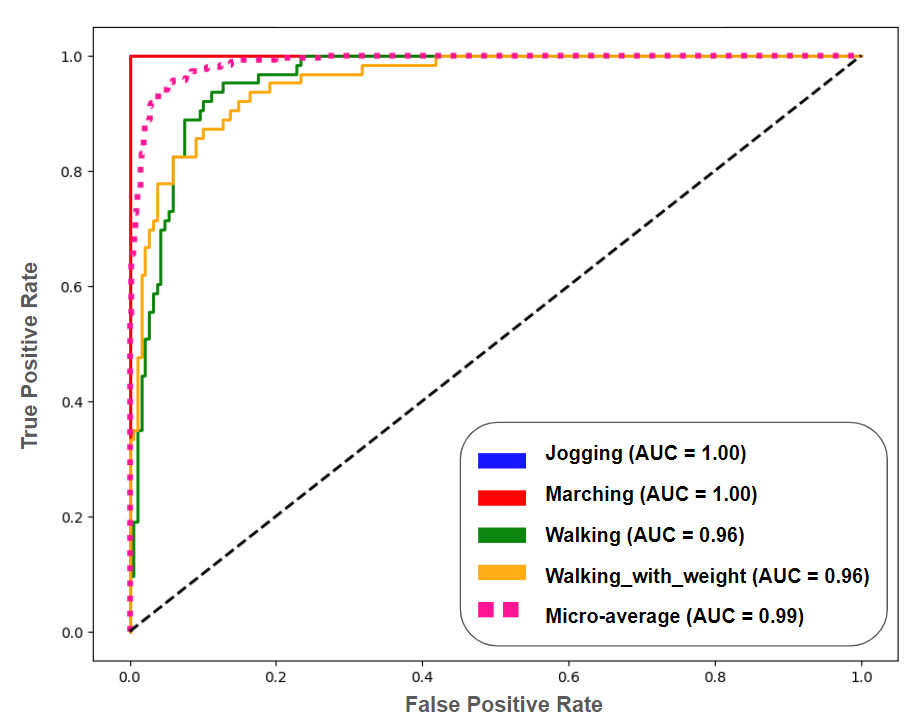}
        \caption{}
    \end{subfigure}
    \hfill
    \begin{subfigure}[b]{0.3\textwidth}
        \includegraphics[width=8.8cm]{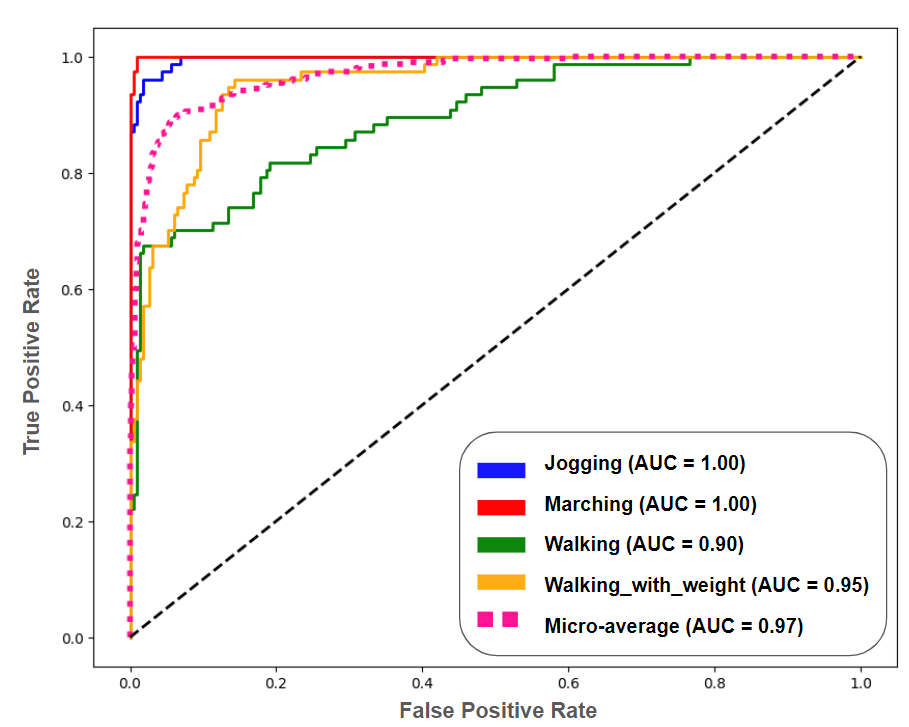}
        \caption{}
    \end{subfigure}
    \caption{The receiver operating characteristics for (a) LSTM model with 1.67s window for magnetic tracking data, and (b) LSTM model with 1.67s window for IMU + magnetometer data.}
    \label{roc}
\end{figure}

We also present the confusion matrices for the run closest to the mean accuracy in Fig \ref{conf_mat}. It can be seen that jogging and marching are activities that are typically classified with high accuracy. There is insignificant performance difference between the magnetic tracking and IMU + magnetometer data. This is because jogging and marching are very easy to distinguish, irrespective of the sensor modality. The gait information content is really put to the test when classifying walking and walking with weight, as they are largely similar looking. We get about 84.45\% average accuracy using magnetic tracking data and LSTM model, as compared to 76.51\% average accuracy using IMU + magnetometer. Thus, we see close to \textbf{8\%} performance improvement when using magnetic tracking data. This result is corroborated by the receiver operating curves (ROC) in Fig \ref{roc}, comparing magnetic tracking and IMU + magnetometer systems. Jogging and marching have areas under the curve (AUC) of 1. Here as well, it can be seen that magnetic tracking system performs better than IMU + magnetometer system in classifying walking and walking with weight activities. 

\begin{figure} [h]
    \centering
    \includegraphics[width=9cm]{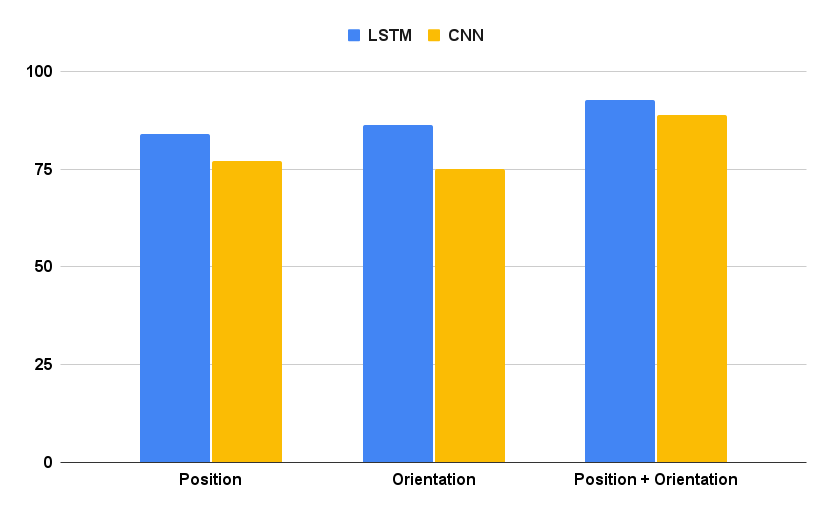}
    \caption{Accuracy comparison of ablation study - position, orientation and position + orientation - using CNN and LSTM models.}
    \label{ablation}
\end{figure}

To study the individual contribution of the position and orientation gait data to the classification performance, an ablation study was conducted using the LSTM model. The same hyperparameters were used for all sensor modalities - position, orientation, position+orientation - for LSTM. For CNN however, due to the change in the input vector dimension for the case of position and orientation, maxpool kernel sizes were reduced to (5,1) and (9,1) from (5,2) and (9,2). The performance comparison for the ablation study is shown in Fig \ref{ablation}. The orientation data has a slightly higher contribution to for classifying the activities compared to position, which could vary depending on the application. Orientation data alone has been used for gait analysis in the past \cite{pierleoni_validation_2019} with good results. We too get a decent accuracy of about 86\% when using LSTM and orientation data. However, combining it with position data shows clear improvement. 

\section{Discussion}
\label{discussion}
Current state-of-the-art gait analysis systems are non-wearable systems \cite{harris_survey_2022}\cite{godinho_systematic_2016}. Non-wearable systems do not have any constraints of power, cost, space, compute power or size. This allows these systems to be sophisticated enough to obtain a holistic information about human gait. However, such systems are not scalable and are not accessible to everyone. Also, in medical applications, observing the gait of patients makes them conscious, which interferes with the analysis. This is called the observer effect or Hawthorne effect \cite{cicirelli_human_2022}. These factors drive research for the betterment of wearable gait analysis systems. Wearable systems are scalable and can potentially capture the gait in the field, with minimal interference. The most popular sensing methods for wearable systems are IMUs, magnetometers and pressure sensors. However, these methods have limitations in how much information related to gait they can provide. Countering this shortcoming by increasing the number of sensors placed on the body comes at the cost of becoming intrusive. As a result, they have not been able to match the performance level of non-wearable systems, making the latter the preferred choice for gait analysis as yet \cite{beauchet_gait_2008}\cite{prasanth_wearable_2021}.

To bridge this gap, we applied our previously developed magnetic tracking system to the purpose of gait analysis. A Human Activity Recognition system is demonstrated as a proof-of-concept. The magnetic tracking system shows better performance compared to the IMU + magnetometer system - on average 4\% better across all ML models and segmentation windows. The better performance of the magnetic tracking system is attributed to the fact that the 6DoF tracking data has more gait information content than the data from IMU and magnetometer. We also see that the LSTM model shows better accuracy than CNN model. LSTM is better able to extract the temporal parameters that are important for gait classification because it is able to capture relationship even between samples that are well separated in time, unlike CNN. Experimenting with other sophisticated ML methods to further improve the classification performance is left for future work.

Differentiating between the activities - marching and jogging - is not that difficult as the gait data generated looks very obviously different. The IMU + magnetometer system also shows good accuracy in differentiating between these 2 activities. However, it has a significantly lower accuracy than the magnetic tracking system in trying to differentiate between walking and walking with weight. These 2 activities were intentionally chosen because the change in gait due to a heavy backpack is very subtle and is a good test for the efficacy of the sensing modality for gait analysis. Being able to differentiate between such subtle gait differences is vital. For example, in the medical field, the initial stages of a neurological condition or a minor injury would cause very subtle changes in the gait. This might be difficult to detect even for an expert. By the time there is a detectable change in gait, the condition or the injury would have worsened. In such scenarios, a gait analysis system that is able to pick up on the subtleties in gait changes would be extremely important. 

One of the main limitations of magnetic tracking systems is the effect of magnetic field distortions due to the presence of metals in the vicinity. However, there has been work on detecting and rejecting this distortion \cite{yadav_accurate_2014}\cite{fan_how_2018}. Whenever such distortion is detected, the tracking system could discard the data, keeping only the legitimate tracking data for gait analysis. Despite this limitation, 6DoF magnetic tracking presents great potential in gait analysis, as demonstrated by this work. We envision the system to be developed into a product like this: The Tx module could be integrated into a belt that could be worn. It could communicate the tracking data via USB to a smartphone that is in the pocket as the person goes about his/her daily activities, with little interference from the system. The data could be transmitted to cloud for performing training and inference, or the inference could be performed on the smartphone itself. The Tx communication to the smartphone could even be done using Bluetooth. This would make the system completely wireless and further reduce its intrusiveness, expanding the scope to other applications such as sports training. The Rx tracker modules can be embedded into the shoe or can be modularized to be simply strapped onto any location on the body. We plan to increase the number of trackers to further improve the gait information, and also experiment with different locations for the trackers, for future work. The data communication pipeline with the smartphone and cloud needs to be set up such that gait data is collected without the user getting interrupted. The tracking system could also be combined with other sensor modalities such as pressure sensors. We have only demonstrated the tracking system for HAR. In future work, the magnetic tracking system should be tested for other gait analysis applications. 

\section{Conclusion}
In this work, we present a wearable magnetic tracking system as a sensing method that can be used for gait analysis. The system is compact, portable and minimally intrusive. It can complement and address the shortcomings of the existing wearable sensing methods, such as, IMUs and pressure sensors. A Human Activity Recognition application was demonstrated using the magnetic tracking system. The position and orientation of both the feet was tracked with respect to a reference point on the waist. Two DL classifiers - CNN and LSTM - were compared. We also compared our system with an IMU + magnetometer system by attaching them to both the feet of participants. The magnetic tracking system shows superior performance to the IMU + magnetometer system. This highlights the insufficiency in the information content in the data from IMUs and magnetometers to capture the complete human gait. Being able to more effectively capture the gait using wearables bridges the gap between the superior performance of non-wearable systems and the portability advantage of wearables. In the presented magnetic tracking system, the number of tracker modules can be increased easily to further improve the capture of gait, which has been planned for the future. We also need to experiment with the location of the tracker modules and combine it with other sensor modalities to improve performance. The authors note that magnetic systems tend to be affected by metals in the vicinity, leading to tracking inaccuracies. Future work should involve developing techniques to detect distortions in the magnetic field to reject the tracking data.

\section*{Acknowledgment}
We thank our friends and acquaintances for generously spending time to volunteer for the gait data collection.

\section{References}
\bibliographystyle{IEEEtran}
\bibliography{IEEE_J-BHI_Gait_Analysis_6DoF_Mag_Track}

\end{document}